\documentclass[preprint,onecolumn,aps,nofootinbib]{revtex4}
\usepackage{graphicx}
\pdfoutput=1
\usepackage[colorlinks=true,linkcolor=blue,urlcolor=blue,filecolor=black,citecolor=red,pdfstartview=FitV,
pdftitle={},pdfsubject={},pdfkeywords={},pdfpagemode=None,bookmarksopen=true]{hyperref}
\usepackage{epsfig}
\usepackage{amsmath}
\usepackage{amsfonts}
\usepackage{amssymb}
\usepackage{color}%
\usepackage{dcolumn}
\providecommand{\U}[1]{\protect\rule{.1in}{.1in}}

\newcommand{\f}{\begin{equation}}
\newcommand{\ff}{\end{equation}}
\newcommand{\fa}{\begin{eqnarray}}
\newcommand{\ffa}{\end{eqnarray}}

\allowdisplaybreaks[4]
\begin{document}
\title{Characterization of Quantum Phase Transition using Holographic Entanglement Entropy}
\author{Yi Ling $^{1,3}$}
\email{lingy@ihep.ac.cn}
\author{Peng Liu $^{1}$}
\email{liup51@ihep.ac.cn}
\author{Jian-Pin Wu $^{2,3}$}
\email{jianpinwu@mail.bnu.edu.cn}
\affiliation{$^1$ Institute of High Energy Physics, Chinese Academy of Sciences, Beijing 100049, China\ \\
$^2$ Institute of Gravitation and Cosmology, Department of
Physics,
School of Mathematics and Physics, Bohai University, Jinzhou 121013, China\ \\
$^3$ Shanghai Key Laboratory of High Temperature Superconductors,
Shanghai, 200444, China}
\begin{abstract}
The entanglement exhibits extremal or singular behavior near
quantum critical points (QCPs) in many condensed matter models.
These intriguing phenomena, however, still call for a widely
accepted  understanding. In this letter we study this issue in
holographic framework. We investigate the connection between the
holographic entanglement entropy (HEE) and the quantum phase
transition (QPT) in a lattice-deformed Einstein-Maxwell-Dilaton theory.
Novel backgrounds exhibiting metal-insulator transitions
(MIT) have been constructed in which both metallic phase and
insulating phase have vanishing entropy density in zero
temperature limit.
We find that the first order derivative of HEE
with respect to lattice parameters exhibits extremal behavior near QCPs.
We propose that it would be a universal feature that HEE or its
derivatives with respect to system parameters can characterize QPT
in a generic holographic system. Our work opens a window for
understanding the relation between entanglement and the QPT from
holographic perspective.
\end{abstract}
\maketitle

\section{Introduction}
The relation between the entanglement and quantum phase transition
(QPT) has extensively been investigated in the field of condensed
matter physics (A recent review on this progress can be found for
instance in \cite{Amico:2007ag}). In particular, it has been
proposed that the entanglement may be used to diagnose the QPT
because it exhibits extremal or singular behavior near the quantum
critical point (QCP) in many models
\cite{Osterloh:2002na,Osborne:2002zz,Vidal:2002rm,YChen:2006jop,Chen:2006eqp,Hamma:2007eft,Anfossi:2006tvm}.
Especially, as an important quantity measuring non-local
correlation, the entanglement entropy (EE) or its first derivative
characterizes the QPT with local extremes near the QCPs
\cite{YChen:2006jop,Chen:2006eqp}. Theoretically, any entanglement
measure can be written as a functional of the first derivatives of
the ground state energy
\cite{Wu:2006leq,Wu:2004prl,Larsson2005eso}, which probably
provides more insight into this issue since QPT is associated with
the drastic modification of the ground state
\cite{Sachdev:2000qpt}. Nevertheless, a widely accepted
understanding on how and why the entanglement characterizes the
QPTs is still missing.

Usually QPT involves strong correlation physics and is hard to
analyze; in quantum field theory EE is also difficult to compute.
Gauge/Gravity duality \cite{Maldacena:1997,Witten:1998} as a novel
approach has provided powerful tools for both
understanding strongly correlated systems and computing EE,
which stimulates us to study the deep relation between
entanglement and QPT from holographic perspective.

Recently, the metal-insulator transition (MIT), as a prominent
example of QPT has been implemented in holographic approach
\cite{Donos:2012js,Donos:2013eha,Donos:2014uba,Ling:2015epa,Ling:2015exa,Kiritsis:2015oxa,Donos:2014oha,Baggioli:2014roa}.
We expect the holographic description of EE
\cite{Ryu:2006bv,Takayanagi:2012kg}, namely the holographic
entanglement entropy (HEE), will also furnish a dramatic signature
of QPT in holographic models. Motivated by this, a connection
between the HEE and QCPs of the MIT has firstly been revealed in
\cite{Ling:2015dma}, which is reminiscent of the relation between
EE and QCP found in CMT \cite{YChen:2006jop}. Nonetheless, in
\cite{Ling:2015dma} the ground state entropy density is vanishing
for insulating phase, while nonvanishing for metallic phase
since its near horizon geometry is $AdS_2\times \mathbb R^2$.
The artifact of $AdS_2$ is a well-known open problem in
AdS/CMT duality since any ordinary condensed matter system in
reality has vanishing entropy density in ground states, otherwise
the system would suffer from the instability.
Hence in \cite{Ling:2015dma} a crucial issue involved is whether the
relation between HEE and QPT would originate from the artifact of $AdS_2$,
and finally prevent this relation from linking to realistic condensed matter systems.
To provide a definite answer to this issue, one direct way is to construct
holographic models with MIT in the absence of $AdS_2$.
This is a challenging task since in all previous holographic realization of MIT, the
instability of $AdS_2$ characterized by its BF bound plays an
essential role in generating a transition to new IR fixed points.
Therefore, to implement MIT from a metallic phase with vanishing
entropy density to an insulating phase, novel mechanism of
inducing instability is needed rather than violating BF bound of $AdS_2$.
In this letter we will construct a novel holographic model which exhibits
a MIT, and both the metallic phase and the insulating phase have vanishing
ground state entropy density.
Remarkably, we find that the location of QCPs can be captured by
the local maximum of the first-order derivative of HEE with
respect to system parameters. Combined with \cite{Ling:2015dma},
the phenomena in CMT that EE or its first order derivatives
\cite{YChen:2006jop,Chen:2006eqp} characterizes the QPT are
reproduced in holographic framework. Next we turn to the
construction of our holographic model.

\section{Holographic Setup}

We start with an action in the framework of
Einstein-Maxwell-Dilaton theory, which contains two $U(1)$ gauge
fields $A$ and $B$, one dilaton field $\Phi_1$ and an additional complex
scalar field $\Phi$. The Lagrangian reads,
\begin{equation}\label{action}
\begin{aligned}
  \mathcal{L} =& R - {\partial _a}\Phi _{}^*{\partial ^a}{\Phi _{}} - \frac{{{m^2}}}{{{L^2}}}|{\Phi _{}}{|^2}
   -\frac{1}{4}G^2- \frac{{{e^{{\Phi _1}}}}}{4}{F^2} \\  & - \frac{3}{2}{\partial _a}{\Phi _1}{\partial ^a}{\Phi _1}
   + \frac{6}{{{L^2}}}\text{Cosh}\left( {{\Phi _1}} \right),
\end{aligned}
\end{equation}
where $F=dA$, $G=dB$ are curvatures of two gauge fields and the
complex field $\Phi$ simulates the Q-lattice structure \cite{Donos:2013eha,Donos:2014uba}.

In the absence of gauge field $B$ and the Q-lattice, this theory
allows a well-known black brane solution with vanishing ground
state entropy density, which now is called Gubser-Rocha solution
\cite{Gubser:2009qt}. In current work the motivation of
introducing Q-lattice structure is twofold. One is to break the
translational symmetry of the background so as to obtain a finite
DC conductivity which has been extensively investigated in
\cite{Donos:2012js,Donos:2013eha,Donos:2014uba,Ling:2015epa,Ling:2015exa,Kiritsis:2015oxa,Donos:2014oha,Baggioli:2014roa,
Horowitz:2012ky,Horowitz:2012gs,Ling:2013nxa,Ling:2014saa,Andrade:2013gsa,Gouteraux:2014hca,Kim:2014bza}.
More importantly, we expect to obtain a MIT which, from the
viewpoint of renormalization group flow, requires that the
Q-lattice deformation must be relevant so that the gravitational
system can run from the original IR fixed point which associates
with a metallic phase, to a new IR fixed point which associates
with an insulating phase. It can be proved that Q-lattices can not
induce such kind of phase transitions over Gubser-Rocha background
in zero temperature limit if only a single gauge field is
present\footnote{We thank Zhuoyu Xian for discussion on this
issue.}. Therefore, we introduce a second gauge field $B$ and
treat it as the Maxwell field. We expect MIT associated with
this gauge field may take place for ground states in this new
framework, and we will justify our expectation in sequent
numerical analysis. Previously this strategy has been adopted in
the construction of holographic Mott-like insulators
\cite{Ling:2015exa}.

We consider the following ansatz for a black brane solution
\begin{eqnarray}\label{metric}
d{s^2} &=&\frac{{{L^2}}}{{{z^2}}}\left( { - fUd{t^2} + \frac{{d{z^2}}}{{fU}} + {V_1}d{x^2} + {V_2}d{y^2}} \right), \nonumber \\
  {A_t}\left(z\right) &=& \frac{\left( {1 - z} \right)}{{1+Qz}}\mu a,\quad  {B_t}\left( z \right)=\frac{\left( {1 - z} \right)}{{1+Qz}}\mu b,\hfill\nonumber\\
  \Phi(z)&=& e^{i\hat kx}z^{3-\Delta}\phi,\quad \hfill {\Phi _1}\left(z\right)= \frac{1}{2}\ln \left( {1 + Qz {\phi _1}}
  \right),
\end{eqnarray}
where $f(z)=(1-z)p(z)/g(z)$ with $p(z)=1+(1+3Q)z +(1+ 3Q(1+Q))z^2$
and $g(z)=(1+Qz)^{3/2}$, while $\mu = L\sqrt{3Q(1+Q)}$. All the
functions $(U,V_1,V_2,a,b,\phi,\phi_1)$ depend on the radial
direction $z$ only. The Hawking temperature of the black brane is
given by $\hat T = 3L\sqrt{1+Q}U(1)/(4\pi )$. When we set
$(U=1,V_1=V_2=g(z),a=1,b=0,\phi=0,\phi_1=1)$, the present ansatz
goes back to Gubser-Rocha solution presented in
\cite{Gubser:2009qt,Ling:2013nxa}. In particular, the Gubser-Rocha
solution in zero temperature limit can be obtained as
$Q\rightarrow \infty$, with dimensionless temperature $T\equiv\hat
T/\mu=\sqrt{3}/(4\pi\sqrt{Q})$, and dimensionless entropy density
$s\equiv\hat s/\mu^2=\sqrt{1+Q}/(3Q)$, thus the entropy density
vanishes linearly with the temperature.

Next we numerically solve the background equations of motion based on the ansatz in Eq.(\ref{metric}).
The requirement of UV region $(z=0)$ being asymptotic $AdS_4$ leads to $V_1(0)=V_2(0)=U(0)=1$ and the
scaling dimension of $\Phi$ is $\Delta = 3/2 + \sqrt{9/4+m^2}$. For definiteness we fix $L=1\,,m^2=-2$, thus $\Delta=2$.
In addition, we set the boundary condition for the rest functions
as $a(0)=1$, $b(0)\equiv b_0$, $\phi(0)=\hat \lambda$ and $\phi_1 (0)=1$.
Within this setting the chemical potential of the dual field corresponding to $A$
is $\mu$ and we will use it as the unit of scaling throughout this letter.
As a result, the family of background solutions based on ansatz (\ref{metric}) can be parameterized by four
dimensionless quantities, namely $\{\hat T/\mu, \,\hat\lambda/\mu,\, \hat k/\mu,\, b_0\}$. For simplicity we
abbreviate these dimensionless quantities as $\{T,\lambda,k,b_0\}$. The parameters $(\lambda,k)$ related to
the Q-lattice $\Phi$ can be interpreted as the lattice strength and as the lattice wave vector.

Without loss of generality we will concentrate on the family of
black brane solutions with fixed $b_0=0.1$. We have also worked
with other values of $b_0$, and qualitatively similar phenomena
can be observed. It is worthwhile to point out that for large
$b_0$ insulating phases exist only for large values of $\lambda$
at which the numerical work will be very costly.

\section{MIT and Ground State Entropy Density}

In this section we investigate the conductivity behavior of the current dual to gauge
field $B$ and the entropy density of the dual system in zero temperature limit.

At zero temperature an insulator has vanishing direct-current
conductivity $\sigma_{DC}$, while a metal has non-vanishing
$\sigma_{DC}$. At finite temperature $T$, we practically
distinguish the metallic phase and insulating phase by
$\sigma_{DC}'(T)<0$ and by $\sigma_{DC}'(T)>0$, therefore the
critical line is identified as $\sigma_{DC}'(T)=0$. In this model
the $\sigma_{DC}$ can be computed with the data on the horizon
\begin{equation}\label{hee}
 \sigma_{DC}=\left.\left(\sqrt{\frac{V_2}{V_1}}+\frac{3Q b^2 \sqrt{V_1 V_2}}{2k^2 (1+Q) \phi}\right)\right|_{z=1}.
\end{equation}

\begin{figure}
  \centering
  \includegraphics[width=6cm]{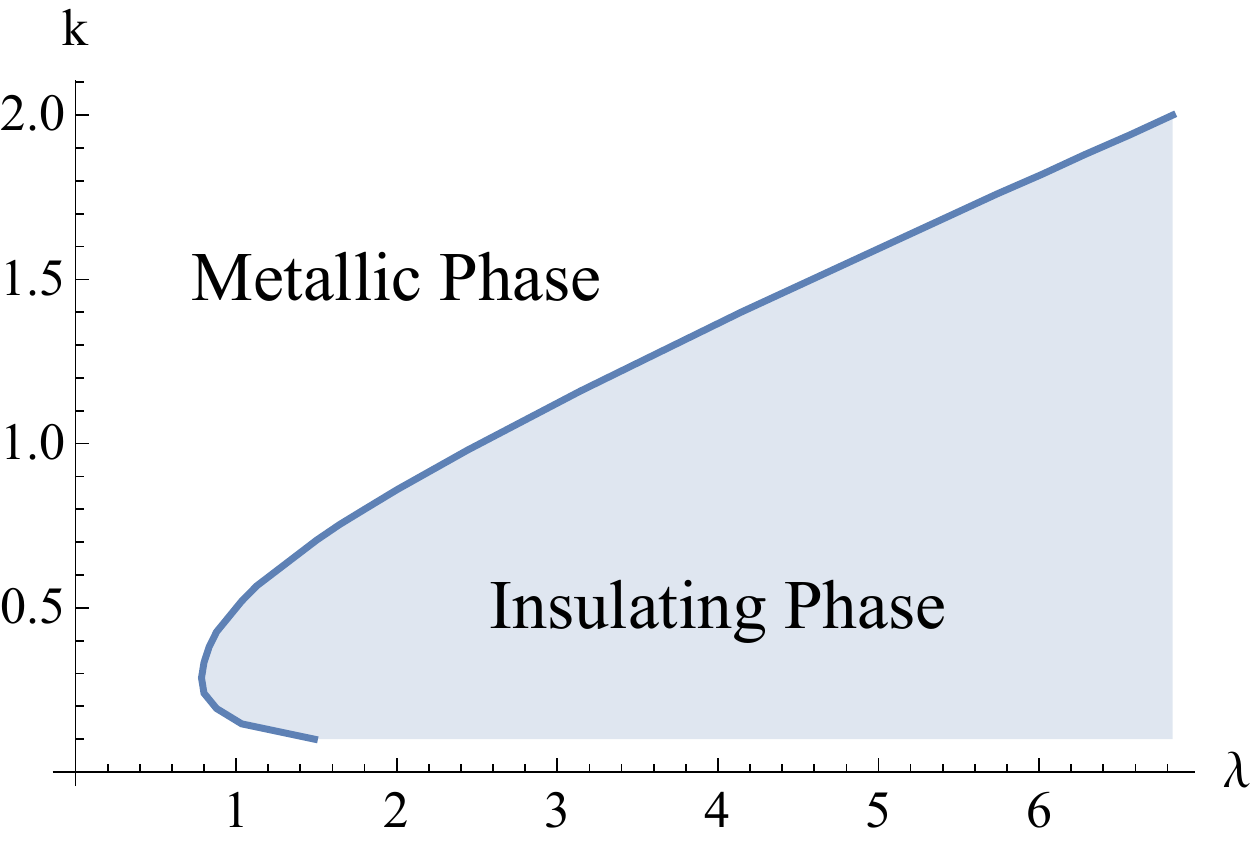}\qquad
   \includegraphics[width=6cm]{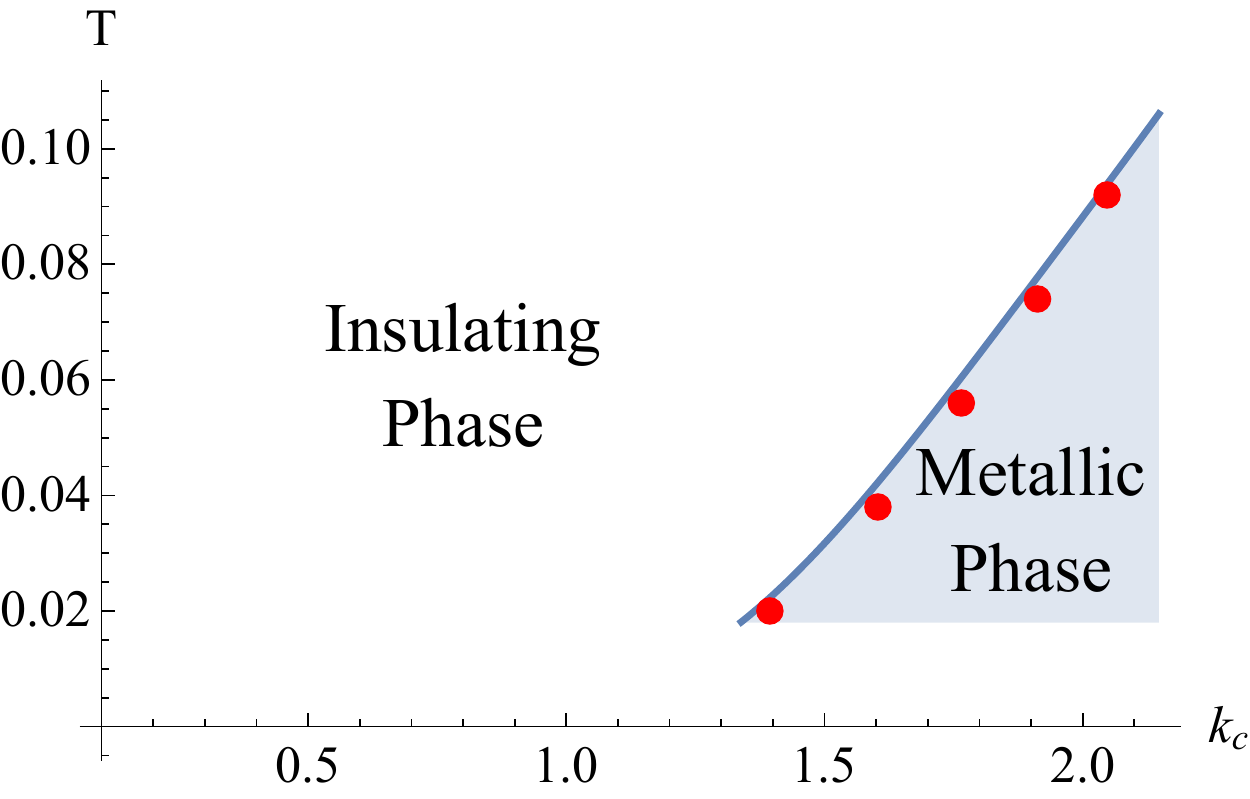}
  \caption{The left plot is the phase diagram over the parameter space $(\lambda, k)$ at $T=0.02$
  with $k> 0.1$. The right plot is the phase diagram over $(T, k)$ plane with $\lambda=4$, $T>0.02$
  and $k>0.1$. The blue curve is the trajectory of the critical points, and the red dots denote the locations
  of the peaks of $\partial_\lambda S(l=6.319)$ at different temperatures when varying the parameter $k$,
  which match well with the critical points of MIT in lower temperature region.}\label{mit}
\end{figure}

First, we demonstrate that a MIT takes place at low temperature as we change the
parameters $(\lambda,k)$ of black brane background.
We plot the phase diagram over the phase space $(\lambda,\,k)$ at $T=0.02$ in Fig.\ref{mit} (the left plot).
In general, it is noticed that the insulating phase is usually formed with large $\lambda$ but small $k$.

Although our phase structure is obtained at a nonzero temperature, we show that a MIT could occur in zero temperature limit.
As illustrated in the right plot of Fig. \ref{mit} (blue curve), $k_c$ changes with the
temperature in an almost linear manner, implying that a non-zero
critical point $k_{c0}$ may be reached at $T=0$.

\begin{figure}
  \centering
  \includegraphics[width=6cm]{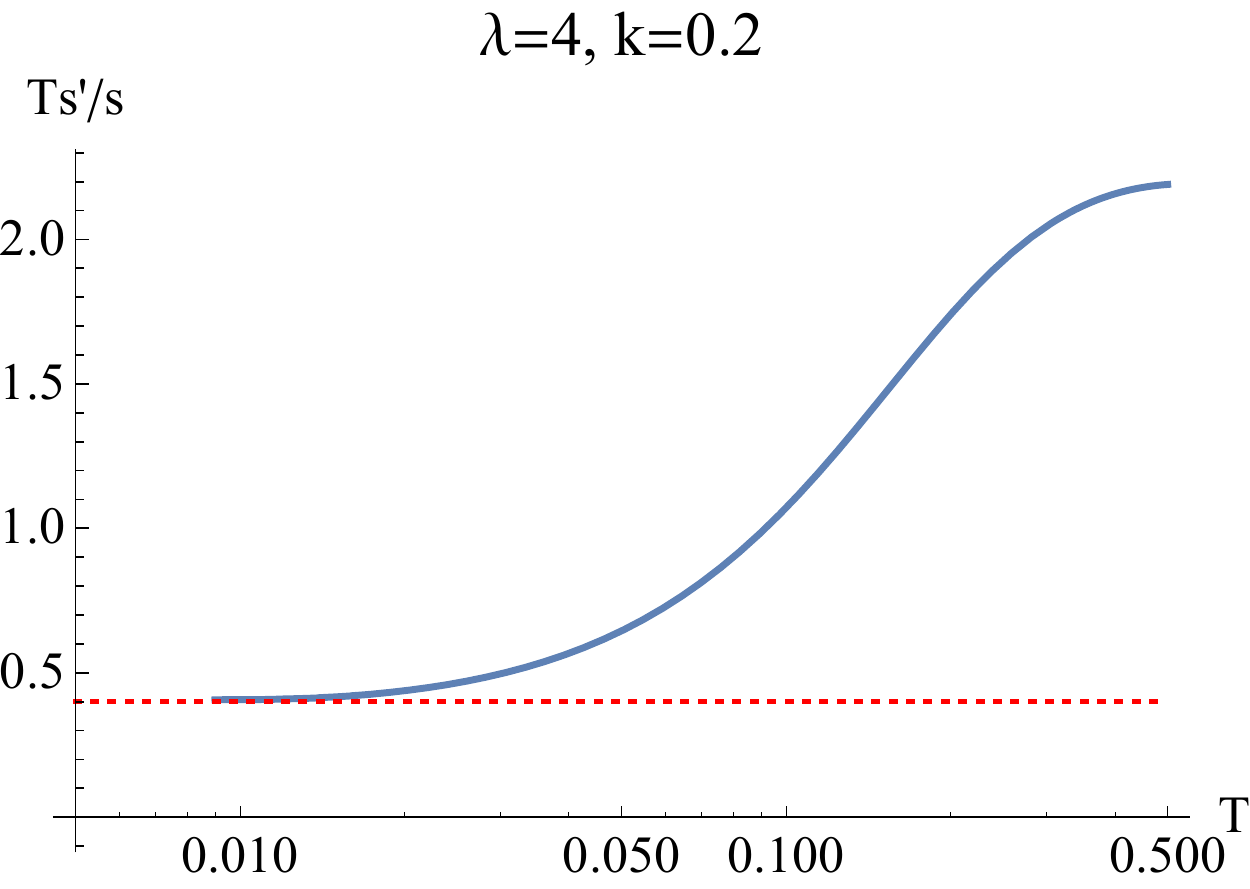}\qquad
  \includegraphics[width=6cm]{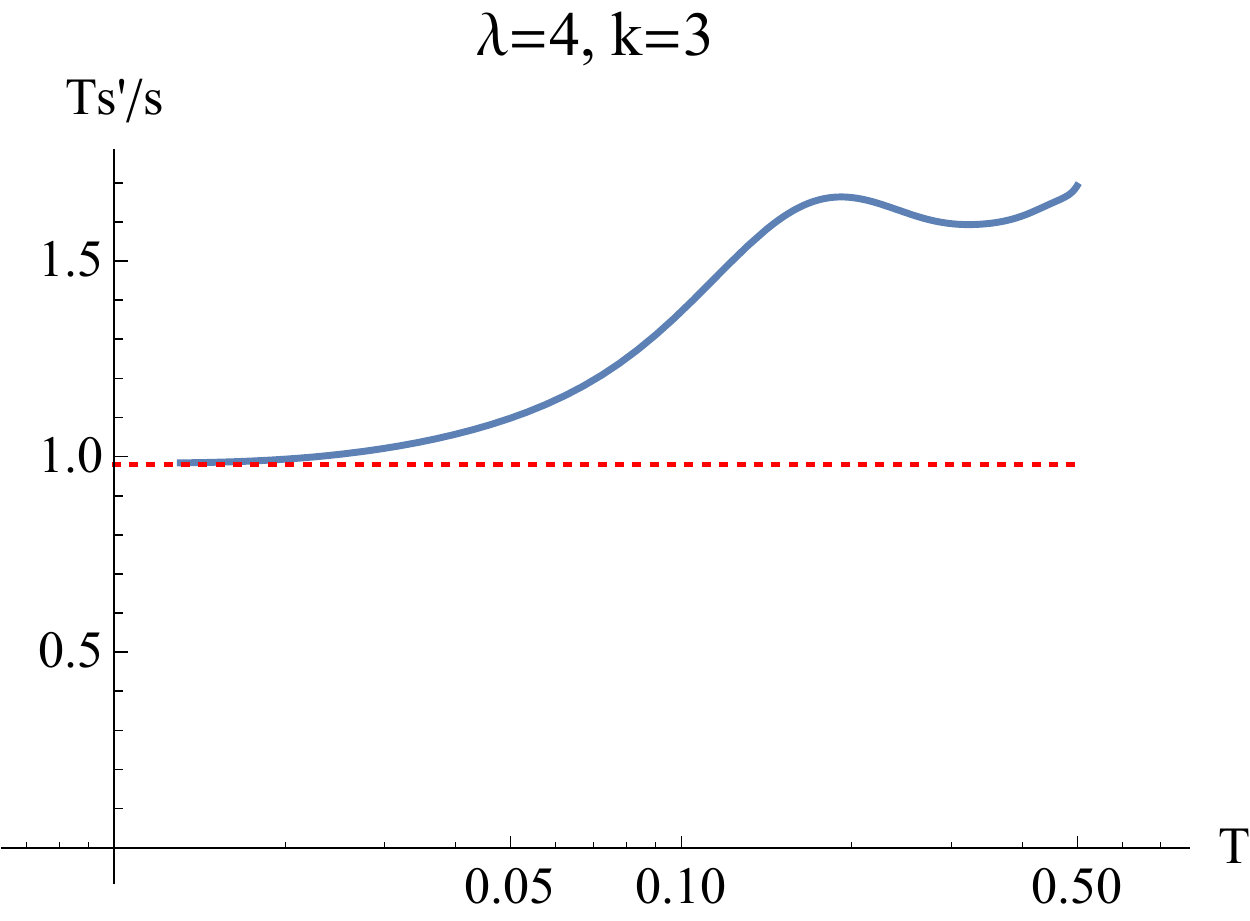}
  \caption{The left and right plots demonstrate the $T\partial_T s/s$ as a function of temperature $T$ in
  insulating phase and metallic phase, respectively.   It can be observed that in metallic phase we have
  $s\sim T^{1}$, and in insulating phase $s\sim T^{0.41}$.}\label{0T0S}
\end{figure}

Now we address the entropy density of black brane in zero temperature limit.
We intend to numerically test the scaling behavior of the entropy density with temperature,
and then justify that our model exhibits vanishing ground state entropy density indeed.
Fig. \ref{0T0S} is a plot of $Ts'/s \; v.s. \; T$ for both metallic phase and insulating phase,
where prime denotes the derivative with respect to $T$. It is clearly seen from this figure that down to an extremely low
temperature $T\sim 10^{-2}$, $Ts'/s$ converges to a
nonzero constant, implying that for both phases the entropy
density has a power law relation with the temperature $s\sim
T^\alpha$. Therefore, the entropy density is expected to be vanishing in zero
temperature limit indeed. Moreover, we notice that the power coefficient
$\alpha$ is always very close to one for metallic phase, while smaller than one for insulating phase,
strongly implying that the RG flow runs to different IR fixed points for a metallic phase
and an insulating phase at zero temperature.

\section{HEE close to QCPs}
In this section, we study the relation between the HEE and QPTs.
In a bipartite system composed of subsystems $A$ and $B$, the EE
of $A$ is defined as $S_A \equiv - \text{Tr}_A \rho_A \ln
(\rho_A)$, where $\rho_A\equiv \text{Tr}_B \rho_\text{total}$.
In AdS/CFT correspondence, the EE for a region $A$ of the boundary system
is obtained from gravity side as the area of the minimal surface
$\gamma_A$ in the bulk which ends at $\partial A$
\cite{Ryu:2006bv,Takayanagi:2012kg}, \emph{i.e.}
\begin{equation}\label{hee}
  S_A=\frac{\text{Area}(\gamma_{A})}{4G_N},
\end{equation}
where $G_{N}$ is the bulk Newton constant.

We consider the HEE for a strip geometry on the boundary with
infinite length $L_y$ in $y$-direction, while with finite width
$\hat l$ in $x$-direction. Notice that the metric components of
the bulk are functions of $z$ only, the minimal surface bounded
with the strip can be specified by the location $z_*$ of the
bottom of the minimal surface in $z$-direction. Therefore, the HEE
$S_A$ could be written as $S_A = L_y\hat S/4G_N$, we will omit the
common prefactor and treat $\hat S$ as the HEE. The
scaling-invariant HEE $S$ and strip width $l$ satisfy the
following equations,
\begin{equation}\label{numhee}
\begin{aligned}
S& =-\frac{2}{\mu z_*} + \frac{2}{\mu} \int_{0}^{z_*}\frac{\chi z_*^2 V_2(z) \sqrt{ V_1(z)}-1}{z^2}, \\
l&=2\mu \int_{0}^{z_*}\chi  z^2\sqrt{\frac{V_1\left(z_*\right) V_2\left(z_*\right)}{V_1(z)}},
\end{aligned}
\end{equation}
where $\chi \equiv \left[{(1-z)p(z)U(z)\xi/g(z)}\right]^{-1/2}$
with $\xi \equiv z_*^4 V_1(z) V_2(z)-z^4 V_1(z_*) V_2(z_*)$,
$S=\hat S/\mu$ and $l=\hat l \mu$. Note that we have subtracted
out the vacuum contribution to HEE by adding a counter term
$-1/z^2$ into the integrant of $S$.
\begin{figure}
  \centering
  \includegraphics[width=9cm]{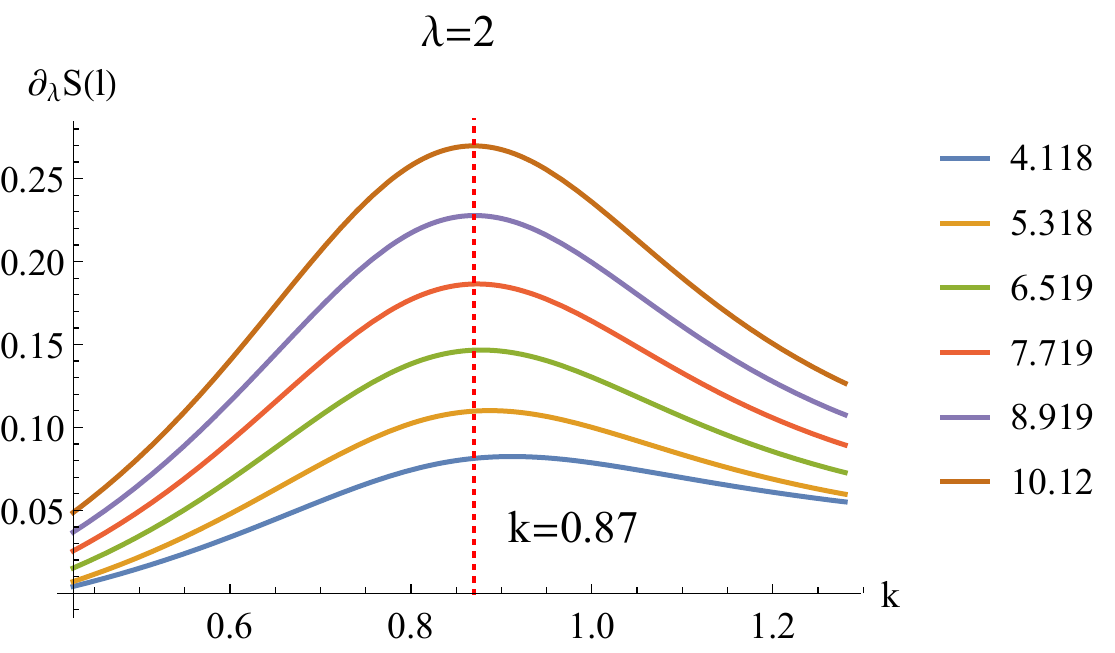}
  \caption{$\partial_\lambda S \; v.s. \; k$ for different values of $l$ with $T=0.02$ and $\lambda=2$.
  It can be seen that for $l \gtrsim 4$,  $\partial_\lambda S$ reaches its local extreme at $k\simeq 0.87$,
  independent of the width of strip.}\label{newp1}
\end{figure}

We study HEE itself as well as the derivatives of HEE with respect to parameters
$(\lambda,k)$. Remarkably, we find in this model it is the first-order derivative of HEE,
$\partial_\lambda S$ and $\partial_k S$, rather than HEE itself that
exhibits an extremal behavior close to critical points. We demonstrate this phenomenon by a
plot $\partial_{\lambda}S \;v.s.\; k$ in Fig. \ref{newp1}. From this figure we notice that the location of peaks is
independent of the width $l$ when it is relatively large, from the
region $Tl\ll 1$ to $Tl\rightarrow \infty$.

Furthermore, we demonstrate that the above phenomena is valid in zero temperature limit.
In the right plot of Fig. \ref{mit} we show the locations of the peaks of $\partial_{\lambda}S$
for different temperatures in the phase diagram $(k,T)$. We find that the peaks always coincide with the critical line
when temperature goes down, indicating that the extremal behavior of the derivative of HEE at
critical points is independent of $T$ in low temperature region. This result allows us to infer safely that this extremal behavior
close to QCPs can be observed in zero temperature limit.

\begin{figure}
  \centering
  \includegraphics[width=9cm]{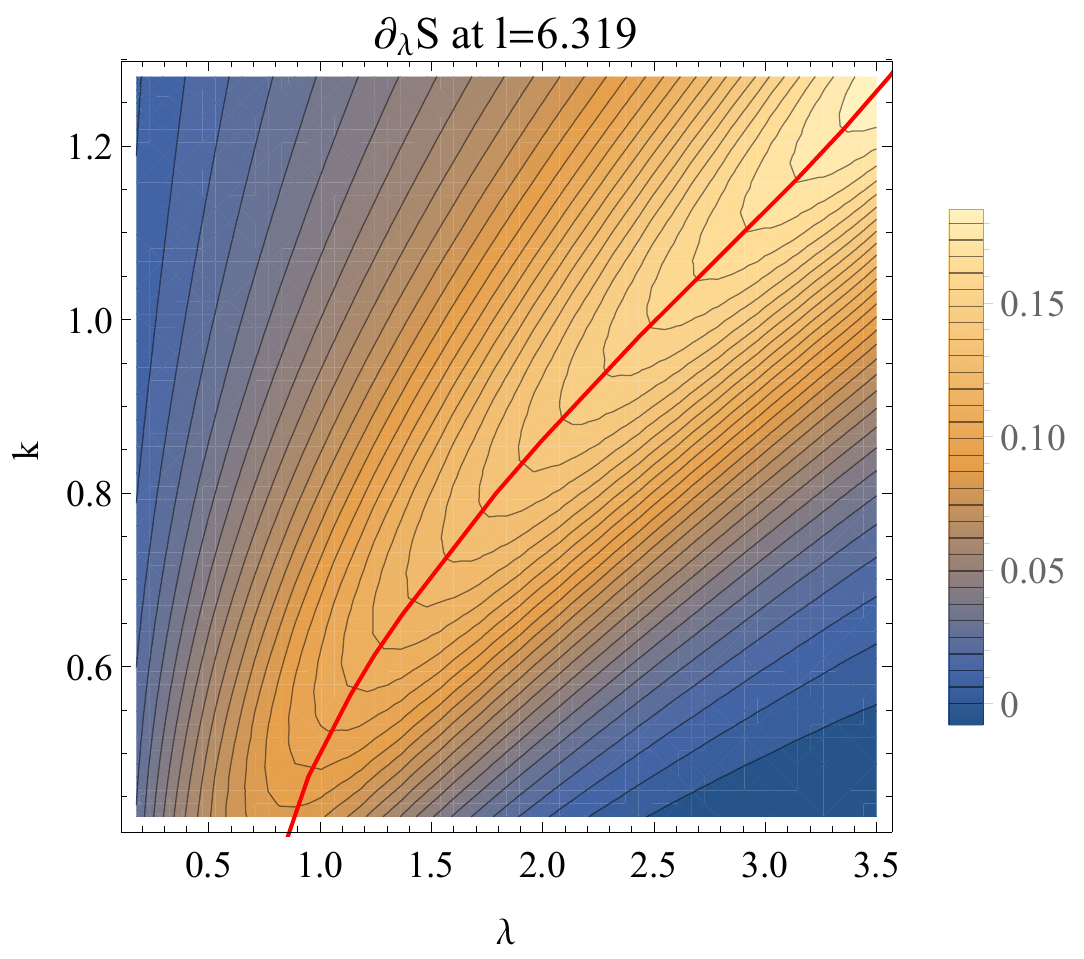}
  \caption{The contour plot of $\partial_\lambda S$ at temperature $T=0.02,\, l=6.319$.
  The red trajectory is the critical line of MIT obtained at $T=0.02$. It can be found that the critical line
  matches well with the ridge of the $\partial_\lambda S$.}\label{contourgrad}
\end{figure}

In parallel, a similar phenomenon can be observed when we change
the parameter $\lambda$ but fix $k$. As a result, we show the
contour plot of $\partial_\lambda S$ over the phase space in Fig.
\ref{contourgrad} with $T=0.02,\, l=6.319$. It can be seen that
the ridge of the contour plot matches very well with the critical
line which characterizes the occurrence of MIT. In addition, we
remark that a similar contour plot can be drawn for $\partial_k
S$, and the only difference is that the ridge is replaced by a
valley.

In summary, we conclude that in this holographic model the first-order derivative of HEE with respect
to system parameters characterizes the QPT.
Previously, the phenomenon that the local extremes of derivatives of EE diagnose
the QPT has been observed in CMT literature \cite{Chen:2006eqp}.
Our holographic model reproduces this significant feature of strongly
correlated system in condensed matter physics.

Finally, we give some remarks on our results in the limit $l\to
\infty$ where HEE is dominated by thermal entropy $S(l)\sim s\cdot
l$. We find that the thermal entropy density $s$ does capture the
locations of critical points as well as $S$ with finite $l$ at low
temperature in our numerical analysis. In zero temperature limit,
however, $s$ vanishes and therefore is not essential in linking
HEE and QPT. As an attempt, it is instructive to define a quantity
$f\equiv S(l)-s\cdot l$ for large $l$ at finite temperature, which
roughly speaking, subtracts out the contribution of pure thermal
entropy $s\cdot l$ from the total HEE $S(l)$. As a matter of fact,
it is shown in \cite{Fischler:2012uv,Kundu:2016dyk} that in the
limit $lT\gg 1$ the $S(l)$ could be separated as thermal
contribution $s\cdot l$ and another finite contribution
$S_{finite}$. The quantity $f$ defined above will become
$S_{finite}$ in large $l$ limit. Numerically, we find that $f$
becomes independent of $l$ for large $l$, and $\partial_\lambda f$
also characterizes the location of QPTs indeed. Moreover, the
ridge of the contour plot of $\partial_\lambda f$ becomes steeper
with the decrease of temperature, and potentially diverges in zero
temperature limit. Therefore, we expect that this quantity would
be helpful for us to investigate the scaling behavior of HEE near
QCPs in zero temperature limit but with $lT\gg 1$. We leave this
for future investigation.

\section{Discussion}

In this letter we have constructed a novel holographic model which
contains black brane solutions with vanishing entropy density in
zero temperature limit, and have demonstrated that a novel
MIT can take place in the absence of $AdS_2$ IR geometry. A
significant improvement to our previous work \cite{Ling:2015dma}
is that an interesting connection between HEE and QPT still exists
in this framework, which means that HEE characterizing QPT
is not an artifact of $AdS_2$. Therefore, our present work
has paved a bridge linking AdS/CMT duality to realistic
condensed matter system. More importantly, it is the first time
to find that the derivatives of HEE with respect to parameters
diagnoses the QPT, which not only coincides with the phenomenon
observed in condensed matter physics, but also enriches our
understanding on the nature of HEE itself. Given the observations
in \cite{Ling:2015dma} and this letter, we conjecture that it
would be a universal feature that the HEE or the derivatives of
HEE could characterize QPT in generic holographic framework. This
conjecture reflects the fact that in CMT the higher-order
derivatives of the measures of entanglement diagnose the QPT
\cite{Larsson:2006sef}.

Our current work has opened a window for understanding the
relation between the HEE and QPT from holographic perspective. It
is intriguing to further investigate the role of HEE in QPT with
the following proposals. First, it is reasonable to expect that
not only HEE or its first order derivative, but also the higher
order derivatives of HEE, can characterize the QPTs in holographic
models. We have gained further evidence to support this conjecture
and the progress will be reported elsewhere \cite{LLWZ}. Second,
it is crucial to understand what determines the order of
derivative of HEE in diagnosing the QPT, which is also an open
question for ordinary EE in condensed matter physics. Further
investigation on this issue should be valuable for disclosing the
nature of both HEE and QPT.

{\it Acknowledgements -} We are very grateful to Chao Niu and
Hongbao Zhang for helpful discussion during the early stage of
this work. This work is supported by the Natural Science
Foundation of China under Grant Nos.11275208, 11305018 and
11575195, and by the grant (No. 14DZ2260700) from the Opening
Project of Shanghai Key Laboratory of High Temperature
Superconductors. Y.L. also acknowledges the support from Jiangxi
young scientists (JingGang Star) program and 555 talent project of
Jiangxi Province. J. P. Wu is also supported by the Program for
Liaoning Excellent Talents in University (No. LJQ2014123).

\end{document}